\documentclass[runningheads]{llncs}
\usepackage[T1]{fontenc}
\usepackage[normalem]{ulem}
\usepackage{utfsym}
\usepackage[misc]{ifsym}
\usepackage{array}
\usepackage{amsmath}
\usepackage[symbol]{footmisc}
\usepackage{cite}
\usepackage[hyphens]{url}
\usepackage{hyperref}
\usepackage[hyphenbreaks]{breakurl}
\useunder{\uline}{\ul}{}
\usepackage{graphicx}
\usepackage{svg}
\begin{document}
\title{Efficient Deep Learning Approaches for Processing Ultra-Widefield Retinal Imaging}
\titlerunning{Efficient UWF Processing}
%
\authorrunning{Kim et al.}

\author{Siwon Kim\inst{1}\thanks{Equal contribution} \and
Wooyung Yun$^{2*}$ \and
Jeongbin Oh\inst{3} \and
Soomok Lee\inst{2}\thanks{Corresponding author}}

\authorrunning{Kim et al.}

\institute{Department of Software, Ajou University, Republic of Korea  \\
\email{kimsiw42@ajou.ac.kr}
\and
Department of Artificial Intelligence, Ajou University, Republic of Korea \\
\email{\{woodolly17, soomoklee\}@ajou.ac.kr}\and
College of Medicine, Seoul National University, Republic of Korea\\
\email{ows0104@snu.ac.kr}}



%
%
%
%
\maketitle

%

\begin{abstract}
Deep learning has emerged as the predominant solution for classifying medical images. We intend to apply these developments to the ultra-widefield (UWF) retinal imaging dataset. Since UWF images can accurately diagnose various retina diseases, it is very important to classify them accurately and prevent them with early treatment. However, processing images manually is time-consuming and labor-intensive, and there are two challenges to automating this process. First, high performance usually requires high computational resources. Artificial intelligence medical technology is better suited for places with limited medical resources, but using high-performance processing units in such environments is challenging. Second, the problem of the accuracy of colour fundus photography (CFP) methods. In general, the UWF method provides more information for retinal diagnosis than the CFP method, but most of the research has been conducted based on the CFP method. Thus, we demonstrate that these problems can be efficiently addressed in low-performance units using methods such as strategic data augmentation and model ensembles, which balance performance and computational resources while utilizing UWF images.
\end{abstract}
\section{Introduction}
Diabetic retinopathy (DR) and diabetic macular edema (DME) are major complications of diabetes and are leading causes of blindness worldwide\cite{Fong2004}. As the number of diabetic patients increases, the incidence of DR and DME are also rising\cite{Stitt2016}. In high-income countries, advanced medical technologies enable effective treatment; however, in low-income countries, diagnostic technologies and treatment resources remain inadequate\cite{Sun2021}. With the global increase in the incidence of diabetes and its complications, establishing effective treatment systems in countries with limited medical infrastructure is essential. This need is not limited to diabetes-related complications; it applies to other diseases as well. Advances in deep learning have facilitated efficient diagnosis even in countries with limited healthcare infrastructure. In environments where the number of patients far exceeds the number of doctors, diagnosing all patients requires a significant amount of time from physicians. To address this issue, researchers have developed various deep learning approaches to assist doctors in diagnostics, many of which have been applied to diabetic retinopathy\cite{Sun2021, Wang2018, Foo2020, park2024fine}.

\begin{figure}[!t]
\includegraphics[width=\textwidth]{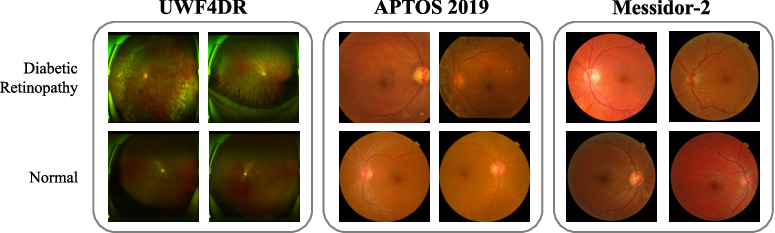}
\caption{Comparison between various retinal imaging datasets. UWF retinal imaging (UWF4DR) and CFP retinal imaging (APTOS 2019, Messidor-2).} \label{fig1}
\end{figure}
However, existing methods have generally been limited to images captured using the widely adopted colour fundus photography (CFP) method\cite{kaggle, Messidor2}, as exemplified in Fig. \ref{fig1}. While CFP is a common retinal imaging methods, it has limitations in identifying peripheral lesions. To overcome these limitations, the ultra-widefield (UWF) retinal imaging method has recently been gaining traction. UWF can capture up to 200 degrees of the retinal periphery, allowing for better identification of peripheral lesions and enabling more accurate diagnoses compared to the CFP method\cite{Silva2016, Sun2016}. Incorporating these advancements in retinal imaging into deep learning applications is crucial. As part of the MICCAI UWF4DR Challenge, we aim to use UWF images in combination with deep learning to develop a more efficient and accurate diagnostic system. 

To classify images using deep learning, existing methods are generally based on either convolutional neural network (CNN)\cite{pratt2016convolutional, raja2022diabetic, zhu2024nnmobilenet} or vision transformer (ViT) architectures\cite{mohan2022vit, nazih2023vision, oulhadj2024diabetic}. CNN-based models effectively learn local features but often struggle to capture important global features. On the other hand, ViT-based models, utilizing the use of self-attention modules, are better suited for capturing global features\cite{karkera2024detecting}. However, ViT models require substantial resources and large datasets to perform effectively. Given that our dataset is relatively small and that many low-income countries face challenges in securing adequate GPU resources, using a ViT-based model is impractical. In such settings, acquiring expensive GPUs is particularly challenging. Therefore, rather than focusing solely on developing a high-performance diagnostic model, we aim to explore methods that enable efficient and fast training even in CPU environments.
For this purpose, we selected the CNN-based EfficientNet\cite{tan2019efficientnet}. EfficientNet is a architecture that promotes balanced scaling of depth, width, and resolution through compound scaling. By using the EfficientNet, along with strategies such as fine-tuning, augmentation, and ensemble techniques, we propose a method that can efficiently train and perform inference with small datasets and limited resources, particularly in low-resource environments. 

\section{Methods}

\begin{figure}[!t]
\includegraphics[width=\textwidth]{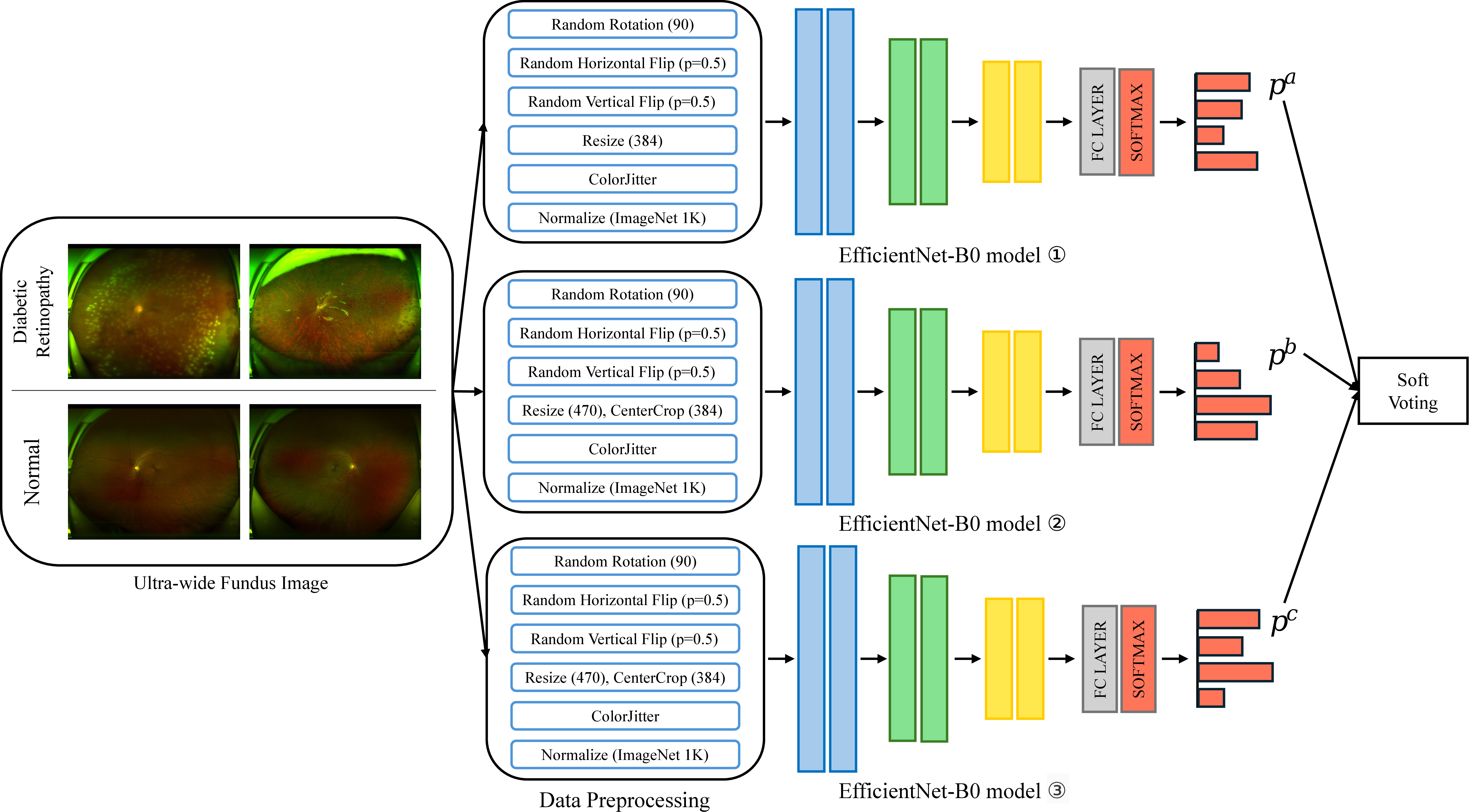}
\caption{The overall pipeline of our proposed method. We utilize EfficientNet-B0 and adopt ensemble learning strategy for robustness.
} \label{fig2}
\end{figure}
In this section, we describe the architecture designed to efficiently detect disease-related features in UWF images. Our proposed method can be seen in Fig. \ref{fig2}.
\subsection{Backbone selection}
In this study, we aimed to use a backbone architecture that can be trained quickly and efficiently even in a CPU environment, without relying on a GPU. To this end, we evaluated the performance of various backbone architectures to select a model that provides both efficiency and high accuracy.

Backbone models can be broadly divided into ViT-based models and CNN-based models. Although ViT-based models have demonstrated excellent performance recently, they are slower in processing and require more resources than CNN-based models, making them more suitable for large-scale datasets. In this study, we selected a CNN-based model, considering the characteristics of the data and the efficiency of the model.

During the model selection process, we experimented with the EfficientNet\cite{tan2019efficientnet}, ResNet\cite{he2016deep}, and ConvNeXt\cite{liu2022convnet} architectures and ultimately selected the EfficientNet. EfficientNet is designed to provide high performance with minimal resources by appropriately scaling the model's depth, width, and resolution\cite{tan2019efficientnet}. In this study, we chose EfficientNet-B0 as the final model, as it uses minimal parameters while still delivering strong performance within the EfficientNet series.

\subsection{Fine-tuning}
We fine-tuned the EfficientNet-B0 model, which was pre-trained on the ImageNet1K dataset, to adapt it to our dataset\cite{deng2009imagenet}. Our proposed method leverages the pre-trained low-level feature information, helping the model converge more stably. Using such a pre-trained model can be advantageous, especially in CPU environments or resource-constrained settings, as it allows for more efficient training.

\subsection{Constant learning rate}
We used the Adam optimizer to update the gradients. The learning rate can be adjusted to control the update speed. Recently various methods have been proposed to gradually reduce the learning rate as the model converges, allowing for more precise approximation\cite{naveen2024cyclical, li2019towards,cutkosky2024mechanic}. However, these methods, which reduce the learning rate over time to help the model reach an optimal convergence point, can also slow down the overall training process.

Given our focus on fast training in a resource-constrained environment, we opted for a fixed learning rate to guarantee a certain level of performance rather than aiming for absolute optimality. This strategy allows us to maintain efficient training while achieving satisfactory results.

\subsection{Data augmentation strategy}
We employed various data augmentation techniques to ensure that the proposed architecture is well-suited for UWF images, preventing overfitting and maintaining robust performance. First, the images were resized to 470 pixels in both width and height, followed by center cropping to 384 pixels. This process effectively removes peripheral areas of the UWF images that are irrelevant to disease diagnosis. While using a smaller resolution could speed up training and reduce resource consumption, we set the resolution to 384 to ensure the extraction of fine details.

Next, we applied random horizontal flip, random vertical flip, and random rotation. These transformations help prevent overfitting and underfitting on retinal images from different orientations, enhancing the robustness of the model. Additionally, we incorporated color jittering by adjusting brightness, contrast, saturation, and hue to ensure that variations in lighting and color during image capture do not adversely affect model performance\cite{zini2022planckian, manuel2022impact}.

These preprocessing techniques significantly aid in accurately detecting lesions in UWF images, allowing the model to detect multiple types of lesions without requiring separate preprocessing for each task.

\subsection{Ensemble strategy}
We employed an ensemble of three EfficientNet-B0 models. The first model was fed with images resized to 384 pixels, while the other two models were given images resized to 470 pixels and center cropped to 384 pixels. By including a model without center cropping, we aimed to capture features that might be lost due to the cropping process, ensuring that regions excluded by center cropping are still detected. Additionally, each model was subjected to different random data augmentations, enhancing robustness and improving the performance of the ensemble.

While the ensemble strategy increases the training time, it is essential for achieving higher performance\cite{dietterich2000ensemble}. For final predictions, we used a soft voting strategy, averaging the probability outputs from the softmax layers of each model to produce the final prediction.

\section{Experiments}
\subsection{Datasets and implementation details}
\subsubsection{UWF Dataset.} UWF dataset provides images captured from a wider field of view compared to the traditional CFP method, enabling more accurate classification of retinal diseases. This dataset is labeled and divided into three tasks. Task 1 involves a binary classification to determine whether the captured image is suitable for analysis. Task 2 distinguishes between patients with referable DR and normal individuals, while Task 3 focuses on diagnosing DME and distinguishing it from normal cases. In this study, we focus on Tasks 2 and 3, which classify diseases in the UWF dataset. The dataset is divided into a training set for model training, a validation set, used to evaluate mid-competition rankings, and a test set for the final evaluation. The validation and test sets are not available to participants and are evaluated via a separate scoring server. The distribution of data for each task is summarized in Table \ref{tab2}.
\begin{table}[!t]
\caption{Data distribution of UWF datasets in each task.}\label{tab2}
\resizebox{\textwidth}{!}
{%
\begin{tabular}{l|cc|cl|cl}
\hline
\textbf{Task}                           & \multicolumn{2}{c|}{\textbf{Train}} & \multicolumn{2}{c|}{\textbf{Validation}} & \multicolumn{2}{c}{\textbf{Test}} \\ \hline
label                                   & \multicolumn{1}{c|}{0}      & 1     & \multicolumn{2}{c|}{-}                   & \multicolumn{2}{c}{-}             \\ \hline
Task 1 - Image Quality Assessment       & \multicolumn{1}{c|}{205}    & 229   & \multicolumn{2}{c|}{61}                  & \multicolumn{2}{c}{Non-public}    \\ 
Task 2 - Referable Diabetic Retinopathy & \multicolumn{1}{c|}{90}     & 112   & \multicolumn{2}{c|}{50}                  & \multicolumn{2}{c}{Non-public}    \\ 
Task 3 - Diabetic Macular Edema         & \multicolumn{1}{c|}{91}     & 77    & \multicolumn{2}{c|}{45}                  & \multicolumn{2}{c}{Non-public}    \\ \hline
\end{tabular}}
\end{table}
\subsubsection{Implementation details.}
In this study, we utilized the Adam optimizer with a fixed learning rate of 0.001. The binary cross-entropy loss function was employed, with a batch size set to 64. Training was conducted for a total of 10 epochs, and the model weights from the epoch with the lowest loss were selected as the final submission weights.

Experiments for backbone selection were performed using the PyTorch framework on a single Intel Xeon Silver 4310 CPU, while the ablation study was conducted in the Codalab evaluation environment.

\subsection{Comparison for backbone selection} We experimented with three CNN-based architectures—EfficientNet\cite{tan2019efficientnet}, ResNet\cite{he2016deep}, and ConvNeXt\cite{liu2022convnet}—to select the best backbone model. The results of these experiments are shown in Table \ref{tab3}. To evaluate both accuracy and efficiency, we compared the area under roc curve(AUROC) and the number of parameters for each model to determine the optimal choice. For all experiments, 50\% of the publicly available training dataset was used for training, while the remaining 50\% was used for validation. The AUROC was calculated based on Task 2, which involves classifying the presence of DR. The experiments were repeated three times with fixed seed values (2023, 2024, 2025), and the average results were used for model selection. Each model was trained for 10 epochs.

While models with more parameters may improve performance over a longer training period, we limited the training to 10 epochs to focus on fast and efficient learning. As a result, models with more parameters were relatively less converged. Among the models tested, EfficientNet-B2 achieved the highest average AUROC, followed by EfficientNet-B0 in terms of performance. However, considering that EfficientNet-B0 has approximately half the number of trainable parameters compared to EfficientNet-B2, we selected EfficientNet-B0 as the final backbone model due to its greater efficiency.

\begin{table}
\caption{Comparison of the AUROC and trainable parameters by backbone model. The best performance is highlighted in bold, and the second is underlined.}\label{tab3}
\resizebox{\textwidth}{!}
{%
\begin{tabular}{c|cccc|cl}
\hline
\textbf{Model}   & \multicolumn{4}{c|}{\textbf{AUROC}}                                                                                                              & \multicolumn{2}{c}{\textbf{Params}} \\ \hline
                 & \multicolumn{1}{c|}{\textbf{Seed(2023)}} & \multicolumn{1}{c|}{\textbf{Seed(2024)}} & \multicolumn{1}{c|}{\textbf{Seed(2025)}} & \textbf{Mean}   & \multicolumn{2}{c}{}                    \\ \hline
ResNet-18\cite{he2016deep}         & \multicolumn{1}{c|}{0.9831}              & \multicolumn{1}{c|}{0.9956}              & \multicolumn{1}{c|}{0.9852}              & 0.9883          & \multicolumn{2}{c}{11.1M}               \\ 
ResNet-34\cite{he2016deep}         & \multicolumn{1}{c|}{0.9651}              & \multicolumn{1}{c|}{1.0000}                 & \multicolumn{1}{c|}{0.9838}              & 0.9829          & \multicolumn{2}{c}{21.1M}               \\ 
ResNet-50\cite{he2016deep}         & \multicolumn{1}{c|}{0.9758}              & \multicolumn{1}{c|}{0.9909}              & \multicolumn{1}{c|}{0.9909}              & 0.9858          & \multicolumn{2}{c}{23.5M}               \\ 
EfficientNet-B0\cite{tan2019efficientnet} & \multicolumn{1}{c|}{0.9971}              & \multicolumn{1}{c|}{0.9952}              & \multicolumn{1}{c|}{0.9960}              & {\ul 0.9961}    & \multicolumn{2}{c}{4.0M}                \\ 
EfficientNet-B1\cite{tan2019efficientnet} & \multicolumn{1}{c|}{0.9950}              & \multicolumn{1}{c|}{0.9964}              & \multicolumn{1}{c|}{0.9960}              & 0.9958          & \multicolumn{2}{c}{6.5M}                \\ 
EfficientNet-B2\cite{tan2019efficientnet} & \multicolumn{1}{c|}{0.9983}              & \multicolumn{1}{c|}{0.9984}              & \multicolumn{1}{c|}{0.9999}              & \textbf{0.9989} & \multicolumn{2}{c}{7.7M}                \\ 
ConvNeXt-tiny\cite{liu2022convnet}   & \multicolumn{1}{c|}{0.9991}              & \multicolumn{1}{c|}{0.8801}              & \multicolumn{1}{c|}{0.7537}              & 0.8776          & \multicolumn{2}{c}{14.9M}               \\ 
ConvNeXt-small\cite{liu2022convnet}  & \multicolumn{1}{c|}{0.9950}              & \multicolumn{1}{c|}{0.8286}              & \multicolumn{1}{c|}{0.8297}              & 0.8844          & \multicolumn{2}{c}{49.4M}               \\ \hline
\end{tabular}
}
\end{table}

\subsection{Ablation study}
\subsubsection{Augmentation.} In this study, we conducted an extensive experiment to validate the effectiveness of each data augmentation technique we applied. To ensure that the improvements in performance were not limited to a specific task, but were applicable to diagnosing various diseases in UWF images using the same architecture, we performed experiments on both Task 2 and Task 3. The AUROC and CPU time were calculated using the test dataset within the Codalab evaluation environment\cite{codalab}. CPU time was measured as the average inference time per image on the Codalab environment CPU.

The experimental results, shown in Table \ref{tab4}, demonstrate that applying all of our proposed augmentation techniques led to higher AUROC scores for both Task 2 and Task 3. This confirms that these augmentation methods significantly contributed to improving the model's performance on the UWF dataset.

\begin{table}
\caption{Ablation study of various augmentation techniques. The best performance is highlighted in bold, and the second is underlined. 
CC, RR, RF, and CJ denote centercrop, random rotation, random flip, and color jittering.}\label{tab4}

\resizebox{\textwidth}{!}
{%
\begin{tabular}{c|c|c|c|c|c|c|c|c}
\hline
\textbf{Model}   & \textbf{CC, Resize} & \textbf{RR} & \textbf{RF} & \textbf{CJ} & \textbf{\begin{tabular}[c]{@{}c@{}}Task 2\\ AUROC\end{tabular}} & \textbf{\begin{tabular}[c]{@{}c@{}}Task 2\\ CPU Time\end{tabular}} & \textbf{\begin{tabular}[c]{@{}c@{}}Task 3\\ AUROC\end{tabular}} & \textbf{\begin{tabular}[c]{@{}c@{}}Task 3\\ CPU Time\end{tabular}} \\ \hline
EfficientNet-B0\cite{tan2019efficientnet} & \textbf{	\usym{1F5F8}}                                                           & \textbf{	\usym{2715}}               & \textbf{	\usym{2715}}           & \textbf{	\usym{2715}}            & 0.8953                                                          & 0.0508                                                             & 0.8852                                                          & 0.048                                                              \\ 
EfficientNet-B0\cite{tan2019efficientnet} & \textbf{	\usym{1F5F8}}                                                           & \textbf{	\usym{1F5F8}}               & \textbf{	\usym{2715}}           & \textbf{	\usym{2715}}            & 0.9271                                                          & 0.0462                                                             & 0.9059                                                          & 0.0486                                                             \\ 
EfficientNet-B0\cite{tan2019efficientnet} & \textbf{	\usym{1F5F8}}                                                           & \textbf{	\usym{1F5F8}}               & \textbf{	\usym{1F5F8}}           & \textbf{	\usym{2715}}            & {\ul 0.9380}                                                    & 0.0459                                                             & \textbf{0.9281}                                                 & 0.0465                                                             \\ 
EfficientNet-B0\cite{tan2019efficientnet} & \textbf{	\usym{1F5F8}}                                                           & \textbf{	\usym{1F5F8}}               & \textbf{	\usym{1F5F8}}           & \textbf{	\usym{1F5F8}}            & \textbf{0.9390}                                                 & 0.0536                                                             & {\ul 0.9278}                                                    & 0.0463                                                             \\ \hline
\end{tabular}
}
\end{table}

\subsubsection{Ensemble.} In this study, we conducted an ablation study to evaluate the effectiveness of the ensemble strategy proposed in the Method section, which involves combining models utilizing three different augmentation techniques. The experimental results shown in Table \ref{tab5}, indicate that increasing the number of high-performing models in the ensemble leads to higher AUROC. However, we also observed that while performance improves with ensemble techniques, there is a corresponding increase in inference time proportional to the number of models used in the ensemble.

\begin{table}
\caption{Ablation study to check the ensemble results of the proposed architecture.}\label{tab5}
\resizebox{\textwidth}{!}
{%
\begin{tabular}{c|c|c|c|c|c|c}
\hline
\textbf{\begin{tabular}[c]{@{}c@{}}Ensemble\\ Model 1\end{tabular}} & \textbf{\begin{tabular}[c]{@{}c@{}}Ensemble\\ Model 2\end{tabular}} & \textbf{\begin{tabular}[c]{@{}c@{}}Ensemble\\ Model 3\end{tabular}} & \textbf{\begin{tabular}[c]{@{}c@{}}Task 2\\ AUROC\end{tabular}} & \textbf{\begin{tabular}[c]{@{}c@{}}Task 2\\ CPU Time\end{tabular}} & \textbf{\begin{tabular}[c]{@{}c@{}}Task 3\\ AUROC\end{tabular}} & \textbf{\begin{tabular}[c]{@{}c@{}}Task 3\\ CPU Time\end{tabular}} \\ \hline
\textbf{\usym{1F5F8}}                                                          & \textbf{\usym{2715}}                                                          & \textbf{\usym{2715}}                                                          & 0.9380                                                          & 0.0536                                                             & 0.9278                                                          & 0.0463                                                             \\ 
\textbf{\usym{2715}}                                                          & \textbf{\usym{1F5F8}}                                                          & \textbf{\usym{2715}}                                                          & 0.9322                                                          & 0.0480                                                             & 0.9316                                                          & 0.0521                                                             \\ 
\textbf{\usym{2715}}                                                          & \textbf{\usym{2715}}                                                          & \textbf{\usym{1F5F8}}                                                          & 0.9349                                                          & 0.0564                                                             & 0.9151                                                          & 0.0562                                                             \\ 
\textbf{\usym{1F5F8}}                                                          & \textbf{\usym{1F5F8}}                                                          & \textbf{\usym{2715}}                                                          & 0.9519                                                          & 0.1005                                                             & 0.9507                                                          & 0.0847                                                             \\ 
\textbf{\usym{1F5F8}}                                                          & \textbf{\usym{2715}}                                                          & \textbf{\usym{1F5F8}}                                                          & 0.9519                                                          & 0.0927                                                             & 0.9383                                                          & 0.1049                                                             \\ 
\textbf{\usym{2715}}                                                          & \textbf{\usym{1F5F8}}                                                          & \textbf{\usym{1F5F8}}                                                          & 0.9567                                                          & 0.0856                                                             & 0.9440                                                          & 0.1009                                                             \\ 
\textbf{\usym{1F5F8}}                                                          & \textbf{\usym{1F5F8}}                                                          & \textbf{\usym{1F5F8}}                                                          & \textbf{0.9662}                                                 & 0.1359                                                             & \textbf{0.9584}                                                 & 0.1322                                                             \\ \hline
\end{tabular}   
}
\end{table}

\section{Discussion and conclusion}
In this study, we investigated efficient deep learning methods for diagnosing various diseases using UWF images. To this end, we evaluated the architecture based on the labeled presence of DR and DME in UWF images. We explored effective model training methods, utilizing fine-tuning, augmentation, and ensemble techniques, to achieve fast convergence and high performance with a lightweight model that can be trained and inferred even in a CPU environment. As a result, we successfully achieved high accuracy in classifying diseases in UWF images, securing 9th place in the MICCAI UWF4DR 2024 Challenge.

Although this study proposed a model that operates efficiently in resource-constrained environments with fast convergence, we found that using the ensemble strategy to improve AUROC performance led to an increase in inference time. Future research could focus on developing augmentation techniques or backbone models capable of operating efficiently in low-resource environments without dependence on ensemble strategies, which would contribute significantly to advancements in medical technology using deep learning.

\subsubsection{\ackname} This research was supported by a grant of ‘Korea Government Grant Program for Education and Research in Medical AI’ through the Korea Health Industry Development Institute(KHIDI), funded by the Korea government(MOE, MOHW).

%
%
%
\bibliographystyle{splncs04}
\bibliography{splncs05}
\end{document}